\documentclass[11pt]{article}
\usepackage{amsmath}
\usepackage{amsfonts}

\renewcommand{\r}[1]{(\ref{#1})}

\newcommand{\be}{\begin{equation}}
\newcommand{\ee}{\end{equation}}
\newcommand{\barr}{\begin{array}}
\newcommand{\earr}{\end{array}}
\newcommand{\bea}{\begin{eqnarray}}
\newcommand{\eea}{\end{eqnarray}}
\newcommand{\beqa}{\be \begin{array}{rcl}}
\newcommand{\eeqa}{\end{array} \ee}
\newcommand{\nn}{\nonumber}

\newcommand{\et}[1]{e^{\mbox{\small $#1$}}}

\newcommand{\ol}[1]{\overline{#1}}

\newcommand{\dt}{{\cdot}}
\newcommand{\wdg}{{\wedge}}
\newcommand{\crs}{{\times}}

\newcommand{\la}{\langle}
\newcommand{\ra}{\rangle}

\newcommand{\half}{{\textstyle \frac{1}{2}}}
\newcommand{\third}{{\textstyle \frac{1}{3}}}

\newcommand{\qrt}{{\textstyle \frac{1}{4}}}

\newcommand{\alp}{\alpha}

\newcommand{\gam}{\gamma}

\newcommand{\Om}{\Omega}

\newcommand{\cla}{{\mathcal{A}}}

\newcommand{\cld}{{\mathcal{D}}}

\newcommand{\clg}{{\mathcal{G}}}

\newcommand{\cll}{{\mathcal{L}}}

\newcommand{\cln}{{\mathcal{N}}}

\newcommand{\clp}{{\mathcal{P}}}
\newcommand{\clq}{{\mathcal{Q}}}
\newcommand{\clr}{{\mathcal{R}}}
\newcommand{\cls}{{\mathcal{S}}}
\newcommand{\clt}{{\mathcal{T}}}

\newcommand{\clv}{{\mathcal{V}}}
\newcommand{\clw}{{\mathcal{W}}}
\newcommand{\clz}{{\mathcal{Z}}}

\newcommand{\da}{\partial_a}
\newcommand{\db}{\partial_b}
\newcommand{\dc}{\partial_c}
\newcommand{\dd}{\partial_d}

\newcommand{\gi}{\gamma_{1}}
\newcommand{\gj}{\gamma_{2}}
\newcommand{\gk}{\gamma_{3}}
\newcommand{\go}{\gamma_{0}}

\newcommand{\Lrev}{\tilde{L}}

\newcommand{\ho}{\ol{h}}

\newcommand{\grad}{\nabla}

\renewcommand{\da}{\gamma^a}
\renewcommand{\db}{\gamma^b}
\renewcommand{\dc}{\gamma^c}
\renewcommand{\dd}{\gamma^d}

\newcommand{\ea}{\gamma_a}
\newcommand{\eb}{\gamma_b}

\newcommand{\numfrac}[2]{{\textstyle \frac{#1}{#2}}}
 
\renewcommand{\i}{I}

\begin{document}

\newcommand{\R}{\clr}
\newcommand{\bv}{\mathbf{v}}

\title{Quadratic Lagrangians and Topology in Gauge Theory Gravity}
\author{Antony Lewis\thanks{aml1005@cam.ac.uk},$^1$\ \ Chris
Doran\thanks{C.Doran@mrao.cam.ac.uk,
http://www.mrao.cam.ac.uk/$\sim$clifford/},$^1$\ \  and Anthony Lasenby$^1$}
\footnotetext[1]{Astrophysics Group, Cavendish Laboratory, Cambridge, U.K.}
\date{\today}
\maketitle
\begin{abstract}
We consider topological contributions to the action integral in a
gauge theory formulation of gravity.  Two topological invariants are
found and are shown to arise from the scalar and pseudoscalar parts of
a single integral.  Neither of these action integrals contribute to
the classical field equations.  An identity is found for the
invariants that is valid for non-symmetric Riemann tensors,
generalizing the usual GR expression for the topological invariants.
The link with Yang-Mills instantons in Euclidean gravity is also
explored.  Ten independent quadratic terms are constructed from the
Riemann tensor, and the topological invariants reduce these to eight
possible independent terms for a quadratic Lagrangian.  The resulting
field equations for the parity non-violating terms are presented.  Our
derivations of these results are considerably simpler that those found
in the literature. 
\end{abstract}


\section{Introduction}

In the construction of a gravitational field theory there is
considerable freedom in the choice of Lagrangian.  Einstein's theory
is obtained when just the Ricci scalar is used, but there is no
compelling reason to believe that this is anything other than a good
approximation.  Since quadratic terms will be small when the curvature
is small one would expect them to have a small effect at low energies.
However they may have a considerable effect in cosmology or on
singularity formation when the curvature gets larger.  Quadratic terms
may also be necessary to formulate a sensible quantum theory.

In this paper we consider the effects of quadratic Lagrangians when
gravity is considered as a gauge theory.  Topological invariants place
restrictions on the number of independent quadratic terms one can
place in the Lagrangian.  In the gauge theory approach these
invariants arise simply as boundary terms in the action integral.  The
Bianchi identity means that these terms do not contribute to the
classical field equations, though they could become important in a
quantum theory.  The invariants have a natural analog in Euclidean
gravity in the winding numbers of Yang-Mills instantons.  These are
characterized by two integers which can be expressed as integrals
quadratic in the Riemann tensor.

Here we investigate instantons and quadratic Lagrangians in Gauge
Theory Gravity (GTG) as recently formulated by Lasenby, Doran and
Gull~\cite{DGL98-grav}.  GTG is a modernized version of ECKS or $U_4$
spin-torsion theory where gravity corresponds to a combination of
invariance under local Lorentz transformations and diffeomorphisms.
With a Ricci Lagrangian GTG reproduces the results of General
Relativity (GR) for all the standard tests, but also incorporates
torsion in a natural manner.  When quadratic terms are introduced into
the Lagrangian the theories differ markedly.  In GR one obtains fourth
order equations for the metric~\cite{Stelle78}, whereas in GTG one has
a pair of lower order equations.  One of these determines the
connection, which in general will differ from that used in GR.  A
reason for these differences can be seen in the way that the fields
transform under scale transformations.  In the GTG approach, all of
the quadratic terms in the action transform homogeneously under
scalings.  In GR the only terms with this property are those formed
from quadratic combinations of the Weyl tensor.

We start with a brief outline of GTG, employing the notation of the
Spacetime Algebra (STA)~\cite{hes-sta,hes-gc}.  This algebraic system,
based on the Dirac algebra, is very helpful in elucidating the
structure of GTG.  The simplicity of the derivations presented here is
intended in part as an advertisement for the power of the STA.  We
continue by constructing the topological invariants for the GTG action
integral.  We show that the two invariants are the scalar and
pseudoscalar parts of a single quantity, and our derivation treats
them in a unified way.  The relationship with instanton solutions in
Euclidean gravity is explored.  As for instantons in Yang-Mills
theory the rotation gauge field becomes pure gauge at infinity and the
topological invariants are the corresponding winding numbers. 

We construct irreducible fields from the Riemann tensor and use these
to form ten independent quadratic terms from the Riemann tensor.  In
an action integral the two topological terms can be ignored, so only
eight terms are needed.  We construct the field equations for the
parity non-violating Lagrangian terms.  Units with $\hbar=c=8\pi G=1$
are used throughout.

\section{Gauge Theory Gravity (GTG)}

In this paper we employ the Spacetime Algebra (STA), which is the
geometric (or Clifford) algebra of Minkowski spacetime.  For details
of geometric algebra the reader is referred
to~\cite{hes-sta,hes-gc}.  The STA is generated by 4
orthonormal vectors, here denoted $\{\gam_\mu\}, \mu=0\cdots 4$.  These
are equipped with a geometric (Clifford) product.  This product is
associative, and the symmetrized product of two vectors is a scalar:
\begin{equation} 
\half (\gam_\mu \gam_\nu + \gam_\nu \gam_\mu) = \gam_\mu \dt \gam_\nu = \eta_{\mu\nu}
= \mbox{diag}(+---).
\end{equation}
Clearly the $\gam_\mu$ vectors satisfy the same algebraic properties as
the Dirac matrices.  There is no need to introduce an explicit matrix
representation for any of the derivations presented here.  The
antisymmetrized product of two vectors is a bivector, denoted with a
wedge $\wedge$.  For two vectors $u$ and $v$ we therefore have
\begin{equation}
uv = \half(uv+vu) + \half(uv-vu) = u \dt v + u \wdg v.
\end{equation}
These definitions extend to define an algebra with 16 elements:
\begin{equation} 
\begin{array}{ccccc}
1 &  \{ \gam_\mu \} & \{ \gam_\mu \wdg \gam_\nu \} & \{ I \gam_\mu \} & I \\ 
\mbox{1 scalar} & \mbox{4 vectors} & \mbox{6 bivectors} & 
\mbox{4 trivectors} & \mbox{1 pseudoscalar} \\
\mbox{grade 0} & \mbox{grade 1} & \mbox{grade 2} & \mbox{grade 3} &
\mbox{grade 4},
\end{array} 
\label{sta}
\end{equation}
where the pseudoscalar $I$ is defined by
\begin{equation}
I \equiv \go\gi\gj\gk.
\end{equation}
The pseudoscalar satisfies $I^2=-1$, and generates duality
transformations, interchanging grade-$r$ and grade-$(4-r)$
multivectors.

The STA approach to gauge theory gravity, or GTG, was introduced in~\cite{DGL98-grav}.
The notation there relied heavily on the use of geometric calculus.
Here we have chosen to adopt a different notation which is closer to
more familiar systems.  These conventions are sometimes not as elegant as those
of~\cite{DGL98-grav,DGL98-spintor}, but they should help to make the
results more accessible.  The first of the gravitational gauge fields
is a position-dependent linear function mapping vectors to vectors.
In~\cite{DGL98-grav} this was denoted by $\ho(a)$.  Here we will
instead write
\begin{equation}
h^a = \ho(\da)
\end{equation}
The metric $g^{ab}$ can be formed from 
\begin{equation}
g^{ab}=h^a \dt h^b
\end{equation}
Clearly $h^a$ is closely related
to a vierbein and this relationship is explained in detail
in the appendix to~\cite{DGL98-grav}.  One point to note is that only one type of
contraction is used in GTG, which is that of the underlying
STA~\r{sta}.  Our use of Latin indices reflects the fact that these
indices can be read as abstract vectors, and can be regarded as a
shorthand for the notation of~\cite{DGL98-grav,DGL98-spintor}. Of
course the index can also be viewed as a reference to a particular
orthogonal frame vector $\gam_\mu$.

The second gauge field is a bivector-valued function $\Om_a$.
This ensures invariance under local Lorentz transformations, which are
written in the STA using the the double-sided formula
\begin{equation}
A \mapsto L A \Lrev.
\end{equation}
Here $A$ is an arbitrary multivector, $L$ is a \textit{rotor} --- an
even element satisfying $L\Lrev=1$ --- and the tilde denotes the
operation of reversing the order of vectors in any geometric product.
Under a Lorentz transformation $\Om_a$ transforms as
\begin{equation}
\Om_a \rightarrow L\Om_a\Lrev - 2\grad_a L\Lrev,
\end{equation}
where $\grad_a = \gam_a \dt \grad$ is the flat space derivative in the $\gam_a$
direction.  It follows that $\Om_a$ takes its values in the Lie
algebra of the group of rotors, which in the STA is simply the space
of bivectors.  Of course $\Om_a$ is a form of spin
connection, the difference here being that it takes its values
explicitly in the bivector subalgebra of the STA.

The $\Om_a$ function is used to construct a derivative which is
covariant under local spacetime rotations.  Acting on an arbitrary
multivector $A$ we define
\begin{equation}
D_a A \equiv \grad_a A + \Om_a \crs A
\end{equation}
where $\crs$ is the commutator product, $A \times B =
\half(AB-BA)$. 
The commutator of these derivatives defines the field strength,
\begin{equation}
R_{ab} \equiv  \grad_a\Om_b-\grad_b\Om_a + \Om_a\crs\Om_b.
\end{equation}
This is also bivector-valued, and is best viewed as a linear function
of a bivector argument (the argument being $\gam_a\wdg\gam_b$ in this
case).

Note that our notation for the derivative differs slightly from that
in~\cite{DGL98-grav}. We wish to use the convention that fully covariant fields are written in
calligraphic type, so here we use the $\cld_a$ symbol for the fully
covariant derivative
\begin{equation}
\cld_a \equiv \ea\dt h^b D_b.
\end{equation}
We define the covariant field strength, the
\textit{Riemann tensor}, by
\begin{equation}
\R_{ab} \equiv \ea\dt h^c \eb\dt h^d R_{cd}.
\label{Riem}
\end{equation}
Again, $\R_{ab}$ is best viewed as a linear map on the space of
bivectors, and as such it has a total of 36 degrees of freedom.  

These covariant objects are at the heart of the
GTG formalism, and distinguish this approach to one based on
differential forms.  Covariant objects such as $\R_{ab}$, or $\da\cld_a
\alp$ (where $\alp$ is a scalar field), are elements of
neither the tangent nor cotangent spaces.  Instead they belong in a
separate `covariant' space in which all objects transform simply under
displacements.  In this space it is simple to formulate physical laws,
and to isolate gauge invariant variables.

The remaining definition we need is
\begin{equation}
\eb\dt h^a \clt^b  \equiv \db \wdg (\cld_b h^a), 
\end{equation}
which defines the \textit{torsion bivector} $\clt^a$, a covariant
tensor mapping vectors to bivectors.  Since the torsion is not assumed to
vanish, we cannot make any assumptions about the symmetries of
the Riemann tensor.  Specifically the `cyclic identity' of GR,
$\R_{ab}\,\wdg \,\db=0$, no longer holds.

From the Riemann tensor one forms two contractions, the Ricci tensor
$\clr_a$ and the Ricci scalar $\clr$,
\begin{equation}
\R_a = \db \dt \R_{ba} \quad \quad \R=\da\dt\R_a.
\end{equation}
The same symbol is used for the Riemann tensor, Ricci tensor and Ricci
scalar, with the number of subscripts denoting which is intended.
Both of the tensors preserve grade, so it is easy to keep track of the
grade of the objects generated.  The Einstein tensor is derived from
the Ricci tensor in the obvious way,
\begin{equation} 
\clg_a = \clr_a - \half \clr \gam_a.
\end{equation}
These are all of the definitions required to study the role of
quadratic Lagrangians in GTG.

\section{Topological invariants}

We are interested in the behaviour of quadratic terms in the
gravitational Lagrangian in GTG.  We start by constructing the
following quantity (which is motivated by instanton solutions in
Euclidean gravity --- see Section~\ref{S-inst})
\begin{equation}
\clz\equiv \da\wdg \db\wdg \dc\wdg \dd\clr_{cd} \clr_{ab} = \da\wdg
\db\wdg \dc\wdg \dd \half(\clr_{cd} \clr_{ab} +\clr_{ab}\clr_{cd}).
\end{equation}
This is a combination of scalar and pseudoscalar terms only, and so
transforms as a scalar under restricted Lorentz transformations.
From equation~\r{Riem} we can write
\begin{equation}
\clz = h^a\wdg h^b\wdg h^c\wdg h^d R_{cd}R_{ab} 
= h \,\da\wdg \db\wdg \dc\wdg \dd R_{cd}R_{ab} \equiv h Z
\end{equation}
where $h$ is the determinant defined by
\begin{equation}
h^a\wdg h^b\wdg h^c\wdg h^d \equiv h \,\da\wdg \db\wdg \dc\wdg \dd
\end{equation}
and 
\begin{equation}
Z \equiv \da\wdg \db\wdg \dc\wdg \dd R_{cd}R_{ab}.
\end{equation}
We can now form an invariant integral that is independent of the
$h^a$ field as
\begin{equation} 
S\equiv\int |d^4x| h^{-1} \,  \clz = \int |d^4x| Z.
\label{Iint}
\end{equation} 
From the definition of the Riemann tensor we find that
\begin{align}
Z &= \da\wdg\db\wdg\dc\wdg\dd (2\grad_c\Om_d
+\Om_c\Om_d)(2\grad_a\Om_b+\Om_a\Om_b) \nn \\
&=-4\da\wdg\db\wdg\dc\wdg\grad(\grad_c\Om_a\Om_b +
\third\Om_a\Om_b\Om_c) \nn \\ 
&= 2\da\wdg\db\wdg\dc\wdg\grad(R_{ac}\Om_b + \third\Om_a\Om_b\Om_c).
\end{align}
The main step in this derivation is the observation that the totally
antisymmetrized product of 4 bivectors vanishes identically in 4-d.
This proof that $Z$ is a total divergence is considerably simpler than
that given in~\cite{Nieh80}, where gamma matrices were introduced in
order to generate a similar `simple' proof in the Riemann-Cartan
formulation.  Here we have also treated the scalar and pseudoscalar
parts in a single term, which halves the work.

Since the integral reduces to a boundary term it should only
contribute a global topological term to an action integral, and should
not contribute to the local field equations.  This is simple to
check.  There is no dependence on the $h^a$ field, so no
contribution arises when this field is varied.  When the $\Om_a$ field
is varied one picks up terms proportional to 
\begin{equation}
\gam_a\wdg\db\wdg\dc\wdg\dd D_d R_{cb} = \numfrac{1}{3}
\gam_a\wdg\db\wdg\dc\wdg\dd (D_d R_{cb} + D_b R_{dc} + D_c R_{bd}) = 0,
\end{equation}
which vanishes by virtue of the Bianchi identity.  Since the two
topological terms do not contribute to the field equations, and can
therefore be ignored in any classical action integral, it is useful to have
expressions for them in terms of simpler combinations of the Riemann
tensor and its contractions.  For the scalar term (denoted
$\la\clz\ra$) we find that
\begin{align}
\la \clz\ra
&=\da\wdg\db\wdg\dc\wdg\dd\, \R_{cd}\wdg \R_{ab} \nn \\
&=(\da\wdg\db\wdg\dc)\dt[ (\dd\dt\R_{cd})\wdg\R_{ab} +
\R_{cd}\wdg(\dd\dt\R_{ab})] \nn \\
&= (\da\wdg\db)\dt[-\R\R_{ab} + 2\R_c\wdg(\dc\dt\R_{ab}) + \R_{cd}(\dc\wdg\dd)\dt\R_{ab})] \nn \\
&= \R^2+2\da\dt[\db\dt\R_c \, \dc\dt\R_{ab} - \db\dt(\dc\dt\R_{ab})\R_c] +
2 \R_{ba}\dt \bar{\R}^{ab}  \nn \\ 
&= 2 \R_{ba}\dt \bar{\R}^{ab} - 4 \R_a\dt\bar{\R}^a+\R^2,
\end{align}
where the adjoint functions are defined by
\begin{equation}
(\gam_a\wdg\gam_b)\dt\bar{\R}_{cd} \equiv (\gam_c\wdg\gam_d) \dt \R_{ab}
\quad\quad
\gam_a \dt \bar{\R}_b = \gam_b \dt \clr_a.
\end{equation}
For the pseudoscalar term (denoted $\la\clz\ra_4$) we similarly obtain
\begin{align} 
\la\clz\ra_4 
&= \da\wdg\db\wdg\dc\wdg\dd \, \R_{cd} \dt \R_{ab} \nn \\
&= \da\wdg\db\wdg(\bar{\R}_{cd} \, (\dc\wdg\dd) \dt \R_{ab} ) \nn \\
&= -I (\dc\wdg\dd) \dt \R_{ab} \, (I\da\wdg\db)\dt\bar{\R}_{cd} \nn \\
&= 2I \R^*_{cd}\dt\R^{cd}
\end{align}
where we have introduced the dual of the Riemann tensor defined
by
\begin{equation}
\R^*_{ab}\equiv  \half I \ea\wdg \eb\wdg\dd\wdg\dc \R_{cd}.
\end{equation}
We therefore have
\begin{equation} 
S=\int|d^4x| h^{-1} \bigl( 2\R_{ba}\dt\bar{\R}^{ab} -
4\R_a\dt\bar{\R}^a+\R^2  + 2I\R^*_{ab}\dt\R^{ba}\bigr).
\end{equation}
This generalises the usual GR expressions for the topological
invariants to the case where the Riemann tensor need not be symmetric,
as in the case when there is torsion.  Both of the scalar and
pseudoscalar contributions can usually be ignored in the action
integral.  The standard GR expressions are recovered by setting
$\bar{\R}_{ab}=\R_{ab}$ and $\bar{\R}_a = \R_a$.

\section{Relation to Instantons}
\label{S-inst}

The derivation of topological terms in GTG has a Euclidean analog,
which gives rise to instanton winding numbers as found in Yang-Mills
theory.  For this section we assume that we are working in a Euclidean
space.  Most of the formulae go through unchanged, except that now the
pseudoscalar squares to $+1$.  For this section we therefore denote
the pseudoscalar by $E$.  The proof that the integral~\r{Iint} is a
total divergence is unaffected, and so it can be converted to a
surface integral.  The Riemann is assumed to fall off sufficiently
quickly that we can drop the $R_{ac}$ term, so
\begin{equation}
S=-\frac{2}{3}\int|d^3x|n\wdg\da\wdg\db\wdg\dc\,\Om_a\Om_b\Om_c.
\end{equation}
For the Riemann to tend to zero the $\Om_a$ field must tend to pure
gauge,
\begin{equation}
\Om_a=-2\grad_aL\Lrev,
\end{equation}
where $L$ is a (Euclidean) rotor.  The integral is invariant under continuous
transformations of the rotor $L$, so we define the winding numbers
\begin{equation}   
\chi + E\tau \equiv \frac{1}{6\pi^2} \int |d^3x|
n\wdg\da\wdg\db\wdg\dc\, \grad_a L\Lrev\grad_b L\Lrev\grad_c L\Lrev=
\frac{1}{32\pi^2}S.
\label{TopInv}
\end{equation}
The numbers $\tau$ and $\chi$ are instanton numbers for the solution,
here given by the scalar and pseudoscalar parts of one equation.  The
common origin of the invariants is clear, as is the fact that one is a
scalar and one a pseudoscalar.  There are two integer invariants
because the 4-d Euclidean rotor group is $Spin(4)$ and the homotopy
groups obey
\begin{equation}
\pi_3(Spin(4))=\pi_3(SU(2)\crs SU(2)) = \pi_3(SU(2))\crs\pi_3(SU(2)) =
Z\crs Z. 
\end{equation}

In Euclidean 4-d space the pseudoscalar $E$ squares to $+1$ and is used to
separate the bivectors into self-dual and anti-self-dual components,
\begin{equation}
B^{\pm} = \half(1\pm E)B, \quad\quad E B^{\pm} = \pm B^{\pm}.
\end{equation}
These give rise to the two separate instanton numbers, one for each of
the $SU(2)$ subgroups.  In spacetime, however, the pseudoscalar has
negative square and instead gives rise to a natural complex structure.
The structure frequently re-emerges in gravitation theory.  The fact
that the complex structures encountered in GR are geometric in origin
is often forgotten when one attempts a Euclideanized treatment of
gravity.

\section{Quadratic Lagrangians}

We now use the preceding results to construct a set of independent
Lagrangian terms for GTG which are quadratic in the field strength
(Riemann) tensor $\R_{ab}$.  None of these terms contain derivatives
of $h^a$, so all transform homogeneously under rescaling of $h^a$.
Local changes of scale are determined by
\begin{equation} 
h^a \mapsto \et{-\alp} h^a, \quad \Om_a \mapsto \Om_a,
\end{equation}
where $\alp$ is a function of position.  The field strength transforms
as
\begin{equation}
\clr_{ab}\mapsto \et{-2\alp}\clr_{ab},
\end{equation}
so all quadratic terms formed from $\clr_{ab}$ pick up a factor of
$\exp(-4\alp)$ under scale changes.  It follows that all quadratic
combinations contribute a term to the action integral that is
invariant under local rescalings.  This situation is quite different
to GR, where only combinations of the Weyl tensor are invariant.  As
a result the field equations from quadratic GTG (and ECKS theory) are
very different to those obtained in GR.

To construct the independent terms for a quadratic Lagrangian we need
to construct the irreducible parts of the Riemann tensor.  To do this
we write
\begin{equation} 
\R_{ab}=\clw_{ab} + \clp_{ab} +\clq_{ab}
\label{Rdecomp}
\end{equation}
where 
\begin{equation}
\da\clw_{ab}=0 \quad\quad \da\clp_{ab}=\da\wdg\clp_{ab} \quad\quad
\da\dt\clq_{ab}=\R_b.
\end{equation}
In the language of Clifford analysis, this is a form of monogenic
decomposition of $\clr_{ab}$~\cite{ASGDL97,som96}.  To achieve this
decomposition we start by defining~\cite{DGL98-grav} 
\begin{equation}
\clq_{ab}=\half(\R_a\wdg \eb + \ea\wdg\R_b) - \numfrac{1}{6} \ea\wdg
\eb\R,
\end{equation}
which satisfies $\da\dt\clq_{ab}=\R_b$.  We next take the protraction
of~\r{Rdecomp} with $\da$ to obtain
\begin{equation}
\da\wdg\R_{ab} - \half\da\wdg\R_a\wdg \eb = \da\wdg\clp_{ab}.
\label{dawdgr}
\end{equation}
We now define the vector valued function
\begin{equation}
\clv_b \equiv -I\da\wdg\R_{ab} = \da\dt(I\R_{ab}).
\end{equation}
The symmetric part of $\clv_b$ is 
\begin{align}
\clv^+_b 
&=\half(\clv_b + \gam^a \, \clv_a \dt \gam_b) \nn \\
&= -I \half (\gam^a \wdg \clr_{ab} + \gam^a \, \gam_b \wdg \gam^c \wdg
\clr_{ca} ) \nn \\
&= -I(\gam^a \wdg \clr_{ab} -\half \gam^a \wdg \clr_a \wdg \gam_b)
\end{align}
so we have
\begin{equation}
\da\wdg\clp_{ab} = I\clv^+_b.
\end{equation}
It follows that
\begin{equation}
\clp_{ab}= -\half I(\clv^+_a\wdg \eb + \ea\wdg\clv^+_b)+
\numfrac{1}{6} I \ea\wdg \eb \clv
\end{equation}
where
\begin{equation}
\clv=\da\dt\clv_a.
\end{equation}
This construction of $\clp_{ab}$ ensures that the tensor has zero
contraction, as required.

Splitting the Ricci tensor into symmetric and antisymmetric parts we
can finally write the Riemann tensor as
\begin{align}
\R_{ab} =& \clw_{ab} + \half(\R^+_a\wdg \eb + \ea \wdg R^+_b) -
\numfrac{1}{6}\ea\wdg \eb \R \nonumber\\ 
& + \half(\R^-_a\wdg \eb + \ea \wdg R^-_b) - \numfrac{1}{2}I(\clv^+_a\wdg
\eb + \ea\wdg\clv^+_b) +\numfrac{1}{6}I \ea\wdg \eb \clv
\end{align}
where $+$ and $-$ superscripts denote the symmetric and antisymmetric
parts of a tensor respectively.  This decomposition splits the Riemann
tensor into a Weyl term ($\clw_{ab}$) with 10 degrees of freedom, two
symmetric tensors ($\clr^+_a$ and $\clv^+_a$) with 10 degrees of
freedom each, and an anti-symmetric tensor ($\clr^-_a$) with 6 degrees
of freedom.  These account for all 36 degrees of freedom in
$\clr_{ab}$.  The first three terms in the decomposition are the usual
ones for a symmetric Riemann tensor and would be present in GR.  The
remaining terms come from the antisymmetric parts of $\clr_{ab}$ and
only arise in the presence of spin or quadratic terms in the
Lagrangian.  It is now a simple task to  construct traceless
tensors from $\clv^+_a$ and $\R^+_a$ to complete the decomposition
into irreducible parts.

We can write the antisymmetric part of $\R_a$ as
\begin{equation}
\R^-_a = a\dt \cla
\end{equation}
where $\cla=\half \da\wdg\R_a$ is a bivector.  Using this definition
we can write down 10 independent scalar terms which are quadratic in
the Riemann tensor:
\begin{eqnarray}
\quad \{\,\clw^{ab}\dt\clw_{ab}, \quad \clw^{ab}\dt(\i\clw_{ab}), \quad
\R^{+a}\dt\R^+_a, \quad \R^2, \quad \nonumber\\
 \quad \cla \dt \cla, \quad \cla\dt (\i\cla), \quad
\clv^{+a}\dt\clv^+_a,  \quad \clv^{+a}\dt\R^+_a, \quad \clv^2, \quad
\R\clv \,\}
\end{eqnarray}
Six of these are invariant under parity and four are parity violating.
The two topological invariants can be used to remove two terms, so
there are only eight possible independent quadratic terms for the
gravitational Lagrangian.  The classical field equations arising from
an equivalent set of terms is calculated in~\cite{Obukhov89} where the
Einstein-Cartan formalism is used.  The theory is locally equivalent to GTG.

For calculational purposes it is easier to use the six parity
invariant terms
\begin{equation}
\{\,\R^{ab}\dt\R_{ba}, \quad \R^a\dt\R_a, \quad \bar{\R}^a\dt\R_a,
\quad \R^2, \quad \clv^a\dt\clv_a, \quad \clv^2\,\}
\end{equation}
and the four parity violating terms
\begin{equation}
\{\,\R^{ab}\dt(\i\R_{ba}), \quad \R^a\dt\clv_a, \quad
\bar{\R}^a\dt\clv_a,\quad \R\clv\,\}
\end{equation}
which are linear combinations of the irreducible components.  The
topological invariants can be used to remove one term from each set.
If we consider just the parity invariant terms and use the topological
invariant to remove $\bar{\R}^a\dt\R_a$ we can calculate the field
equations from
\begin{equation}
\cll_{R^2}=\qrt\epsilon_1\R^2 +\half\epsilon_2\R^a\dt\R_a
+\qrt\epsilon_3\R^{ab}\dt\R_{ba} +\epsilon_4\qrt\clv^2 +
\epsilon_5\half\clv^a\dt\clv_a 
\end{equation}
The field equations for the $h^a$ give a modified Einstein
tensor of the form
\begin{equation}
\clg'_a = \clg_{a} + \epsilon_1\clg_{1a} + \epsilon_2\clg_{2a} +
\epsilon_3\clg_{3a} + \epsilon_4\clg_{4a} + \epsilon_5\clg_{5a}
\end{equation}
where
\begin{align}
\clg_{1a} &= \R (\R_a - \qrt \ea \R) \\
\clg_{2a} &= \eb\, \R^b\dt\R_a + \R_{ab}\dt\R^b-\half \ea \, \R^b\dt\R_b \\
\clg_{3a} &= \eb\, \R^{bc}\dt\R_{ca}-\qrt \ea \, \R^{bc}\dt\R_{cb}\\
\clg_{4a} &= \clv(\clv_a-\qrt \ea\clv) \\
\clg_{5a} &= \eb\, \clv^b\dt\clv_a + (I\R_{ab})\dt\clv^b-\half \ea\,
\clv^b\dt\clv_b .
\end{align}
These tensors all have zero contraction, as expected from scale
invariance.

The field equations for $\Om_a$ give the generalized torsion equation
of the form
\begin{equation} 
\cln_a=\cls_a
\end{equation}
where $\cln_a$ is the (generalized) torsion tensor and $\cls_a$ is the
matter spin tensor.  Both of these are bivector-valued functions of
their vector argument. 

It is convenient to employ the over-dot notation for the
covariant derivative of tensors,
\begin{equation}
\dot{\cld}_a \dot{T}_b = \cld_a T_b - T_c \, \gam^c \dt (\cld_a \gam_b),
\end{equation}
which has the property of commuting with contractions.  This
definition extends in the obvious manner for tensors with more indices.  The contributions to $\cln_a$ from the five terms in
the action integral are then given concisely by
\begin{align}
\cln_{1a} &= - \clr\, \gam^b \dt ( \gam_a \wdg \clt_b) + \gam_a \wdg \gam^b
\, \cld_b \clr \\
\cln_{2a} &= \bigl( (\gam^b \wdg \gam^c) \dt (\gam_a \wdg \clt_c)
\bigr) \wdg \clr_b + \gam_a \wdg (\dot{\cld}_b \dot{\clr}^b) - \gam^b
\wdg (\dot{\cld}_b \dot{\clr}_a) \\
\cln_{3a} &= \dot{\cld}_b {{\dot{\clr}}_a}{}^{b} + (\gam^b \wdg
\gam^c) \dt \clt_c \, \clr_{ab} - \half \clr_{bc} \, (\gam^c \wdg \gam^b)
\dt \clt_a \\ 
I\cln_{4a} &= - \clv\, \gam^b \dt(\gam_a \wdg \clt_b)  + \, \gam_a \wdg \gam^b \,
\cld_b \clv \\
I\cln_{5a} &= \bigl( (\gam^b \wdg \gam^c) \dt (\gam_a \wdg \clt_c)
\bigr) \wdg \clv_b   + \, \gam_a
\wdg (\dot{\cld}_b \dot{\clv}^b )  - \, \gam^b \wdg (\dot{\cld}_b \dot{\clv}_a). 
\end{align}
More elegant expressions can be obtained if one uses the linear
function notation and conventions used in~\cite{hes-gc,DGL98-grav}.

\section{Conclusions}

We have shown that in gauge theory gravity topological terms are
simply dealt with and reduce to boundary integrals which do not alter
the (classical) field equations.  These topological terms have a
natural analog in the winding numbers for instanton solutions
in Euclidean gravity.  The main difference between the two cases are
due to the opposite signs of the squares of the pseudoscalars.  This
difference is nicely highlighted by working with the scalar and
pseudoscalar invariants in a unified way.  In the Euclidean setup the
pseudoscalar drives duality transformations, which reduce the
$Spin(4)$ group to two $SU(2)$ subgroups.  In Minkowski spacetime,
however, the pseudoscalar has negative square, and is responsible for
the frequently made observation that there is a natural complex
structure associated with the gravitational field
equations~\cite{pen-I}.

We constructed ten possible terms for a quadratic Lagrangian, which
the topological invariants then restrict to eight independent terms.
The field equations for these have been derived elsewhere, but the
derivations and formulae presented here are considerably simpler than
in previous approaches.

\section*{Acknowledgements}
Antony Lewis was supported by a PPARC Research Studentship during the
course of this work.  Chris Doran is supported by an EPSRC Advanced Fellowship.

\end{document}